\documentclass{article}
\input epsf

\def\i{\mbox{i}}
\def\d{\mbox{d}}
\def\Im{\mbox{Im}}
\def\Re{\mbox{Re}}

\begin{document}

\title{\Large \bf Gauge-invariant description of W-pair production\\
in NLO approximation \thanks{To appear in the Proceedings of the
XVth International Workshop ``High Energy Physics and Quantum
Field Theory'', Tver, September 2000 }}

\author{ M. L. Nekrasov \vspace*{2mm}\\ \normalsize Institute for High
Energy Physics \\ \normalsize 142280 Protvino, Russia
\vspace*{-8mm}}
\date{\empty}

\maketitle

\begin{abstract}
\noindent The processes $e^{+}e^{-} \to 4f(\gamma)$ mediated by
W-pair (and single) production are considered in the framework of
the modified perturbation theory, based on the expansion of
probability in powers of the coupling constant instead of
amplitude. A full set of one-loop corrections to cross-section,
and two-loop corrections to self-energy of unstable particles are
completely taken into consideration. It is shown that the
manifestly non-factorizable corrections do not make contributions
into the next-to-leading-order approximation. Moreover, the colour
reconnection does not make contributions within this precision,
either. The rest of the corrections together with the
leading-order contribution provide the gauge-invariant description
of the processes.
\end{abstract}

\bigskip

\noindent $\mbox{\bf 1}$. The current requirement for the
description precision of the processes mediated by production and
decay of W-bosons (as well as Z-bosons) implies the necessity of
direct inclusion of the effects of their instability
\cite{LEP2,NuovoCim,B-P}. A conventional way to take into account
this instability consists in Dyson resummation of self-energies of
unstable particles. This procedure makes it possible to satisfy
the unitarity condition \cite{Veltman} and to avoid the
nonintegrable phase-space singularities that arise while
calculating the probability in the framework of the standard
perturbation theory (PT). However, in the case of pair production
of unstable particles the Dyson resummation violates the Ward
identities (WI) \cite{Argyres,Acta,Dittmaier}. This fact creates a
danger for an uncontrolled error production in the description of
particular processes.

The way to solve this problem consists in making use of the pole
scheme which initially provides the gauge-invariance
\cite{Stuart}. In the case of W-pair production the pole scheme
has been practically applied within the double pole approximation
(DPA) \cite{BBC,RacoonWW,4f}. At present, DPA remains the sole
scheme where the radiative corrections to this process have been
actually evaluated. However, at the Born level DPA turns out to be
untenable, because its direct application leads to the numerical
error of several percent (see, for instance, section 3.5 of
Ref.~\cite{4f}). Therefore, the practical usage of DPA at Born
level is realized with the substitution of the non-gauge-invariant
CC03 off-shell cross-section, instead of the intrinsic
gauge-invariant Born DPA. Thereby it is supposed that this
substitution leads to a reasonable result, at least in the
't$\,$Hooft-Feynman gauge.

The latter substitution may be really numerically reasonable,
at least at LEP energies. Nevertheless, the gauge-invariant
description of the W-pair production, which would be free from the
difficulties of DPA, should be found. Firstly, it should be found
from a conceptual point of view. Secondly, in order to have an
opportunity to estimate the error of the above-mentioned
implementation of DPA. Thirdly, to extend the range of the
description validity. From the viewpoint of future prospects, the
gauge-invariant description should be found as a starting point to
reach the greater precision of the description.

In fact, such gauge-invariant description has already been
proposed \cite{I}. Also a more effective way, in comparison with
DPA, for evaluating the radiative corrections has been found
\cite{I}. The point is that the manifestly non-factorizable
corrections actually do not contribute to the cross-section of
W-pair production within the next-to-leading-order (NLO)
approximation. (Throughout the paper we consider the order of
contributions in the sense of powers of electroweak coupling
constant $\alpha$.) Earlier it was known that the contributions of
the soft-photons to non-factorizable corrections were cancelled in
probability, together with the infrared (IR) divergences
\cite{IR}. However, it remained unknown whether the same property
took place for the intermediate-energy photons. In the DPA
approach the fact of cancellation of the manifestly
non-factorizable corrections to the total cross-section and the
pure angular distributions was confirmed by explicit calculations
\cite{Non}. Nevertheless, the nature of this phenomenon was not
realized. Consequently, it remained unclear how general this
phenomenon was, and whether it occurred in other cases.

The purpose of the present paper is twofold. First, I present the
general proof of the absence of contributions of the manifestly
non-factorizable corrections to the total cross-section in the NLO
approximation, whence the validity of the result in other cases
becomes clear. Then, I propose a further development of the
method. Namely, I show that within the NLO precision the effect of
colour reconnection does not make contribution to the
cross-section of W-pair production.

\medskip

\noindent $\mbox{\bf 2}$. Let us start with discussing the basic
concepts of the approach. Its primary idea \cite{FT} consists in
expanding the probabilities in powers of the coupling constant
instead of amplitudes. Actually this change of operations order
affects the unstable-particle propagators only. Really, while
calculating the probability basing on the completely expanded
amplitude, one faces the divergences in the phase space caused by
the presence of nonintegrable singularities in the free
propagators squared of unstable particles. On the other hand, the
probability is finite in spite of phase-space integration, when it
is calculated on the basis of the (renormalized) amplitude with
the Dyson resummed propagators. Consequently, when the probability
is calculated in this latter way, the result of its expansion
should be a series with finite coefficients.

The latter outcome may be strictly proved by analyzing the
expansion of the Dyson resummed propagator squared. The first term
of this expansion is in fact well-known in view of the following
formula for Breit-Wigner resonance,
\begin{equation}\label{1}
\frac{1}{\left|s\!-\!M^2 + \i M \Gamma \right|^2} \;
\stackrel{\Gamma \to 0}{\longrightarrow} \; \frac{\pi}{ M\Gamma}
\; \delta(s\!-\!M^2) \,.
\end{equation}
Unlike the function $1/|s\!-\!M^2|^2$ that appears as a result of
the expansion before the squaring procedure, the delta-function
$\delta(s\!-\!M^2)$ is integrable with respect to $s$. So, it
produces the finite contribution to the probability while
integrating over the phase space. Obtaining the next terms of the
expansion in formula (1) is a nontrivial problem, but solvable in
the framework of the theory of generalized functions
(distributions) \cite{General} with implements of asymptotic
operation \cite{AO}. Under the assumption of the absence of
contributions of zero-mass particles (photons) the next terms of
the expansion have been derived in \cite{FT}.\footnote{ Although
Ref.~\cite{FT} claimed that it had found some non-trivial
generalization of the method for the case of presence of the
soft-photons, it was a believable result only, not actually
proved, because the actual cancellation of the IR divergences had
not been shown in any physically significant case. The same was
the case with the status of the declared curing \cite{FT} of
difficulties on gauge-invariance.}

Below we present the corresponding result within the NLO
precision. We use the following notation for the {\it
renormalized} self-energy of the unstable particle:
\begin{equation}\label{2}
\Sigma(s;\alpha) = \alpha \Sigma_1(s) + \alpha ^2\Sigma _2(s) +
\,\cdots\;.
\end{equation}
In on-mass-shell (OMS) scheme of UV renormalization (where
$\Re\Sigma_1(M^2)\!=\!\Re\Sigma'_1(M^2)\!=\!0$) the result looks
as follows:
\begin{equation}\label{3}
\frac{1}{\left|s-M^2+\Sigma(s;\alpha)\right|^2} \; = \;
\frac{\pi}{\alpha\Im\Sigma_1(M^2)}\;\delta(s\!-\!M^2) - \pi
\frac{\Im\Sigma_2 (M^2)}{\Im\Sigma_1^2 (M^2)}\;\delta(s\!-\!M^2) +
\mbox{VP}\frac{1}{(s\!-\!M^2)^2} + O(\alpha)\,.
\end{equation}
Here VP means the principal-value prescription for the
nonintegrable in the conventional sense function. Owing to the
unitarity relation $\alpha\Im \Sigma_1(M^2)=M\Gamma$ with $\Gamma$
to be the lowest-order width, which was verified by direct
calculations \cite{Lectures}, the first term in the r.h.s. in (3)
is exactly the r.h.s in formula (1). The next terms in (3)
describe the NLO corrections. It is worth noticing that in the OMS
scheme the real part of the self-energy does not contribute within
the given precision. Moreover, the self-energy contributes in the
on-mass-shell only, and without derivatives. The latter fact means
that the soft-photon problem is not actually present within the
given precision in the propagator squared. However, it arises
while considering the exchanges by soft photons between different
charged particles of the process. Moreover, in the next orders of
the expansion the problem arises even at level of the propagator
squared, raising thereby a question about the validity of the
expansion.

In fact, the problem of IR divergences can be simply solved by
introducing the mass for soft photons. Then, at the end of
calculations it is necessary to show that the soft-photon mass is
cancelled and the gauge invariance is restored. The problem can be
simplified by performing the (secondary) Dyson resummation of the
one-loop corrections. The latter operation is equivalent to an
incomplete expansion in the coupling constant of the full unstable
propagator squared. The general investigation of the problem has
been carried out in \cite{I}. For our purposes the following
resultant formula is relevant:
\begin{equation}\label{4}
\frac{1}{\left|s-M^2+\Sigma(s;\alpha)\right|^2} \; = \;
\frac{1}{\left|s-M^2+\alpha\Sigma_1^{\mbox{\scriptsize fer}}(s)
\right|^2} - \pi \frac{\Im\Sigma_2 (M^2)}{\Im\Sigma_1^2
(M^2)}\;\delta(s\!-\!M^2) + O(\alpha)\,.
\end{equation}
Here $\Sigma_1^{\mbox{\scriptsize fer}}(s)$ is the fermionic
one-loop correction to the self-energy of W-boson. Notice, the
delta-function in the second term in (4) may be regularized by
means of formula (1).

Let us discuss now the properties of formula (4). First of all it
should be noted that its first term is explicitly gauge-invariant.
This fact follows immediately from the explicit gauge-invariance
of $\Sigma_1^{\mbox{\scriptsize fer}}(s)$. Moreover, the first
term in (4) is exactly that quantity which was used in the
framework of fermion-loop scheme \cite{Argyres,Fermion-loop},
proposed for the gauge-invariant approximate description of the
W-pair production. The two-loop corrections, that are important in
the resonance region within the NLO precision
\cite{Acta,Dittmaier} but missed in the fermion-loop scheme, are
collected in the second term in (4). Since this term, originating
in probability, cannot be obtained from the analysis of the
amplitude, it has been called the {\it anomalous} additive
term~\cite{I}.

Actually, the anomalous term is gauge invariant, as well. This
follows from the equality
$\Im\Sigma_1(M^2)=\Im\Sigma_1^{\mbox{\scriptsize fer}}(M^2)$ which
may be derived, in particular, from the corresponding explicit
expressions \cite{B-P}, and this follows from the gauge-invariance
of $\Im\Sigma_2(M^2)$ in the OMS-like schemes \cite{II}. Notice,
the two-loop-level gauge-dependent part in the OMS mass $M^2$
\cite{Sirlin} may be neglected in the anomalous term. Indeed, the
second term in (4) describes the {\it highest}-order correction
within the given precision. So, any higher-order variation in the
mass actually manifests itself as the correction $O(\alpha)$ in
formula (4).

The last but not least property which should be noted, and which
is common for both formulas (3) and (4), is the change of the
natural order of individual contributions from the viewpoint of
the standard PT. Really, the second term in (4) involves a
two-loop correction. However, it describes the NLO correction, but
not the NNLO one. The importance of this property is great~--- it
enables one to selectively take into account only those one-loop
and two-loop corrections which are really needed in the NLO.
Fortunately, the sum of these corrections turns out to be
gauge-invariant. The rest of the one-loop and two-loop
corrections, that contribute beyond the NLO, are gauge-dependent.

\medskip

\noindent $\mbox{\bf 3}$. Let us proceed now to the very
description of the W-pair-mediated four-fermion production in
$e^{+}e^{-}$ annihilation with taking into account the complete
set of NLO corrections. The proposed construction may be presented
as the generalization of the fermion-loop scheme. Remember, the
latter scheme consists in including all fermionic one-loop
corrections in tree-level amplitudes and Dyson resumming the
self-energies. As a result, the amplitude becomes
gauge-independent in spite of the Dyson resummation
\cite{Argyres,Fermion-loop}. However, the fermion-loop scheme does
not include certain corrections that are really important within
the NLO precision \cite{Acta,Dittmaier}. Namely, the fermion-loop
scheme misses the two-loop correction to self-energy in the Dyson
resummed propagators, which is needed in the resonance region, and
all the one-loop bosonic corrections, both to the numerators in
the amplitude and to the denominators in the Dyson resummed
propagators. The idea of the proposed generalization consists in
taking into consideration these corrections in probability but not
in amplitude, i.e. in the framework of the modified~PT.

So, in our approach let the first-step approximation be the
cross-section obtained in the fermion-loop scheme. Actually it
includes completely the leading-order contribution and a part of
the NLO correction. In order to describe another part let us fully
use the property that it is a correction. Consequently, for its
build-up it is enough to know only the leading-order contribution
to the cross-section, but not the full result of the fermion-loop
scheme. It is obvious, that the leading-order contribution
includes the double propagator squared, since only it includes as
far as is possible the lower power of $\alpha$. Namely, it
includes the $\alpha^{-2}$, with singly one $\alpha^{-1}$ twice
arising from the leading-order term in the unstable propagator
squared, cf. formula (3). (Notice, only the transverse parts of
the unstable propagators should be Dyson resummed and squared. The
longitudinal parts of the unstable propagators are to be cancelled
at the end of calculations, by virtue of the expected
gauge-invariance, by contributions of the unphysical states.) Due
to the presence of the delta-function in the leading-order terms
in the unstable propagators squared, the leading-order
contribution to the cross-section should be the cross-section of
the on-shell W-pair production, multiplied by the corresponding
branchings. In fact, this is the cross-section for the CC03
on-shell process.

Let us discuss now the corrections that are missed in the
fermion-loop scheme. As has been mentioned above, there are two
types of these corrections --- the two-loop corrections to
self-energy in the Dyson resummed propagators, and the one-loop
bosonic corrections. The former ones can be taken into
consideration in a quite simple way; one needs only take into
account the anomalous term in formula (4). Since the anomalous
term contributes additively and with the delta-function --- like
the leading-order term in (3) --- its total effect is reduced to
the additive correction of the form of the cross-section of the
on-shell W-pair production multiplied by the corresponding
branchings, and multiplied by the ``anomalous'' factor. The latter
one is determined as follows:
\begin{equation}\label{5}
\delta^{\,\mbox{\scriptsize anom}} =
\left[\frac{\pi}{\alpha\,\Im\Sigma_1(M^2)}\right]^{-1} \times 2
\times \left[-\pi \frac{\Im\Sigma_2 (M^2)}{\Im\Sigma_1^2
(M^2)}\right] = -2\alpha\frac{\Im\Sigma_2 (M^2)}{\Im\Sigma_1
(M^2)}\,.
\end{equation}
Here in the middle expression the first multiplier in square
brackets cancels the coefficient at the delta-function in the
leading-order term in (3), factor 2 is due to the presence of the
two W's in the CC03 process and due to the additive character of
the anomalous term. The last multiplier in (5) is the coefficient
at the delta-function in the anomalous term in (4). Remember, the
anomalous factor is completely gauge invariant due to the gauge
invariance of its ingredients. Actually, without loss of precision
the anomalous factor may be considered as a factor to the full
fermion-loop-scheme cross-section, but not necessary to its
leading-order contribution only, which is the cross-section of the
CC03 on-shell process.

The problem of the one-loop bosonic corrections may be easily
solved, as well. Let us group these corrections into two classes.
To the first class we refer the corrections to self-energies of
unstable particles. To the second class we refer the corrections
to the vertex factors, the corrections due to the exchange
processes, and due to the real (soft) photons.

In fact, the corrections of the first class have already been
taken into consideration by making use of formula (4). Indeed, in
the OMS scheme the bosonic one-loop correction to the W-boson
self-energy does not contribute to the propagator squared within
the NLO precision. This is because only the imaginary part of the
one-loop on-shell self-energy in the OMS scheme is relevant within
the NLO precision. However, the bosonic part of this quantity is
zero.

The bosonic corrections of the second class have not yet been
taken into consideration. To do that, let us make use of the fact
that within the NLO approximation and in the presence of the {\it
corrections}, the Dyson resummed unstable propagators squared may
be considered in the leading-order approximation only. As a
result, the bosonic corrections of the second class may be taken
into consideration as the corrections to the CC03 on-shell
process. Examples of the diagrams of unitarity which generate
contributions of this class are shown in Fig.1. In the presence of
the {\it finite} non-zero mass for the photons all these diagrams
include exactly {\it two} pairs of unstable propagators of the
identical mass and momenta in both sides of the cut. (The only
exception, the configuration of Fig.1i, is discussed below.)
Therefore, they include {\it two} delta-functions of the
corresponding kinematic variables, which make these configurations
on-shell. The sum of all such configurations may be represented as
the sum of two groups of terms. The first group includes the
one-loop cross-section of the pair on-shell production
(Fig.1a,c,e), or the tree-level cross section of the W-pair plus
one real photon on-shell production (Fig.1b,d,f), times the
lowest-order decay blocks of W-bosons. Another group includes the
Born cross-section of the pair on-shell production times their
one-loop decay blocks or the tree-level decay blocks with one real
photon (Fig.\ref1g,h,j).

It should be noted that among the diagrams of Fig.1 there are
configurations of both factorizable and non-factorizable types.
Nevertheless, at an intermediate step, when the photons are
supplied with the finite mass, the non-factorizable configurations
of Fig.1 become factorizable. All other non-factorizable
corrections that do not possess this property, produce
configurations that contain {\it less} than two pairs of the
unstable propagators squared of identical mass and momenta in both
sides of the unitarity cut. Therefore, they do not include the
leading-order factor $1\!\left/ \alpha^2 \right.$. Examples of
configurations of this type are shown in Fig.2. Owing to the
presence of an additional factor $\alpha$ which is due to the
bosonic exchange, all they contribute beyond the NLO. Really, the
configurations of Fig.2a-f contain one propagator squared only,
which includes the factor $1\!\left/ \alpha \right.$.
Consequently, they contribute into the NNLO. The configurations of
Fig.2g,h do not at all contain the propagators squared.
Consequently, they contribute into NNNLO.

Let us note, that the behaviour in $\alpha$ of each
above-discussed configuration does not depend on the value of
photon-mass. So, this behaviour in $\alpha$ should be the same
after the proceeding to the zero-mass limit in the sum of the
corresponding groups of diagrams. The zero-mass-limit operation
can always be done in view of the continuity property of the
above-mentioned cross-sections and decay-blocks, considered as
functions of the photon-mass. Notice, after proceeding to the
zero-mass limit the property of the gauge-invariance must be
restored, if it has been broken by the photon-mass. The
above-mentioned properties of the continuity on the photon-mass
and of the gauge-invariance after the limiting procedure follow
from the well-known theorems for the standard on-shell
cross-sections with contributions of the real soft-photons.

Now let us discuss the above-mentioned exceptional configuration
of Fig.1i. Strictly speaking, it should not be considered among
the diagrams of unitarity, since it describes the self-energy
correction to the unstable propagator, which has already been
taken into consideration in formula (4). Nevertheless, while
considering the cross-section of the pair on-shell production, one
has to take into account the virtual soft-photon insertions to the
external legs, which are due to the wave function renormalization.
The configuration of Fig.1i was added to the list of diagrams of
Fig.1 only in order to indicate this fact.

\begin{figure*} \hbox{
\hspace*{-5pt}
       \epsfxsize=77pt \epsfbox{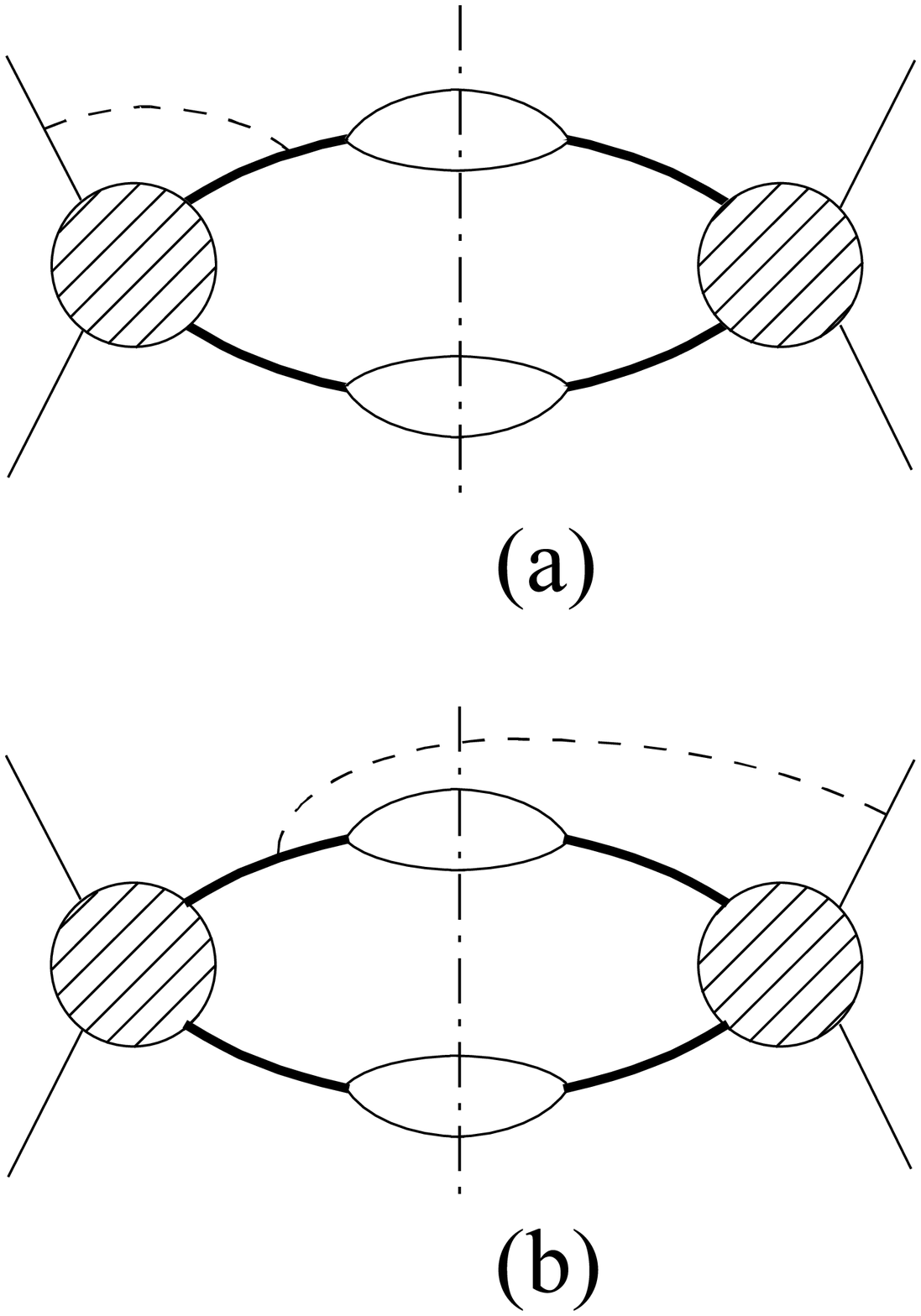} \hspace*{10pt}
       \epsfxsize=77pt \epsfbox{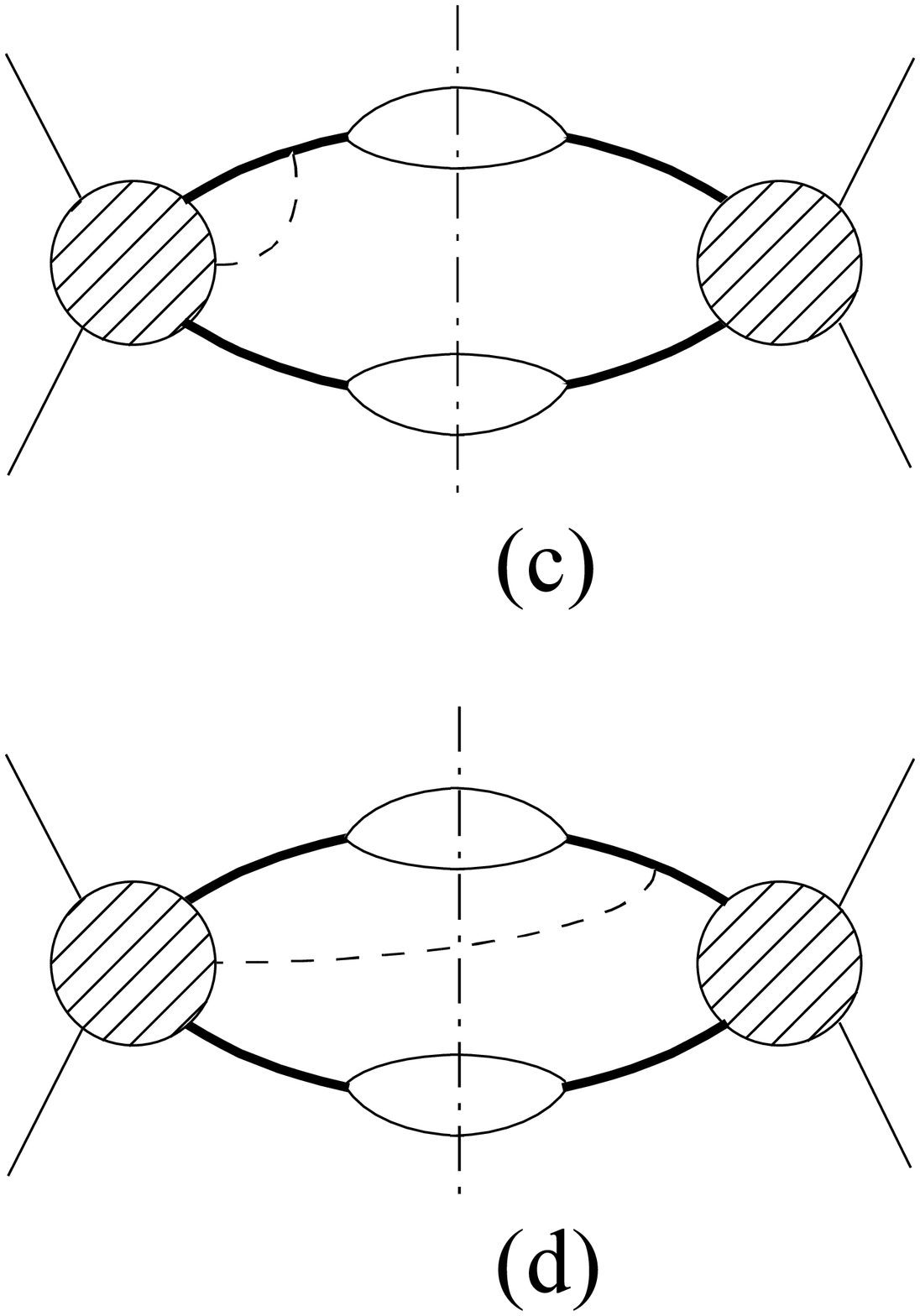} \hspace*{10pt}
       \epsfxsize=77pt \epsfbox{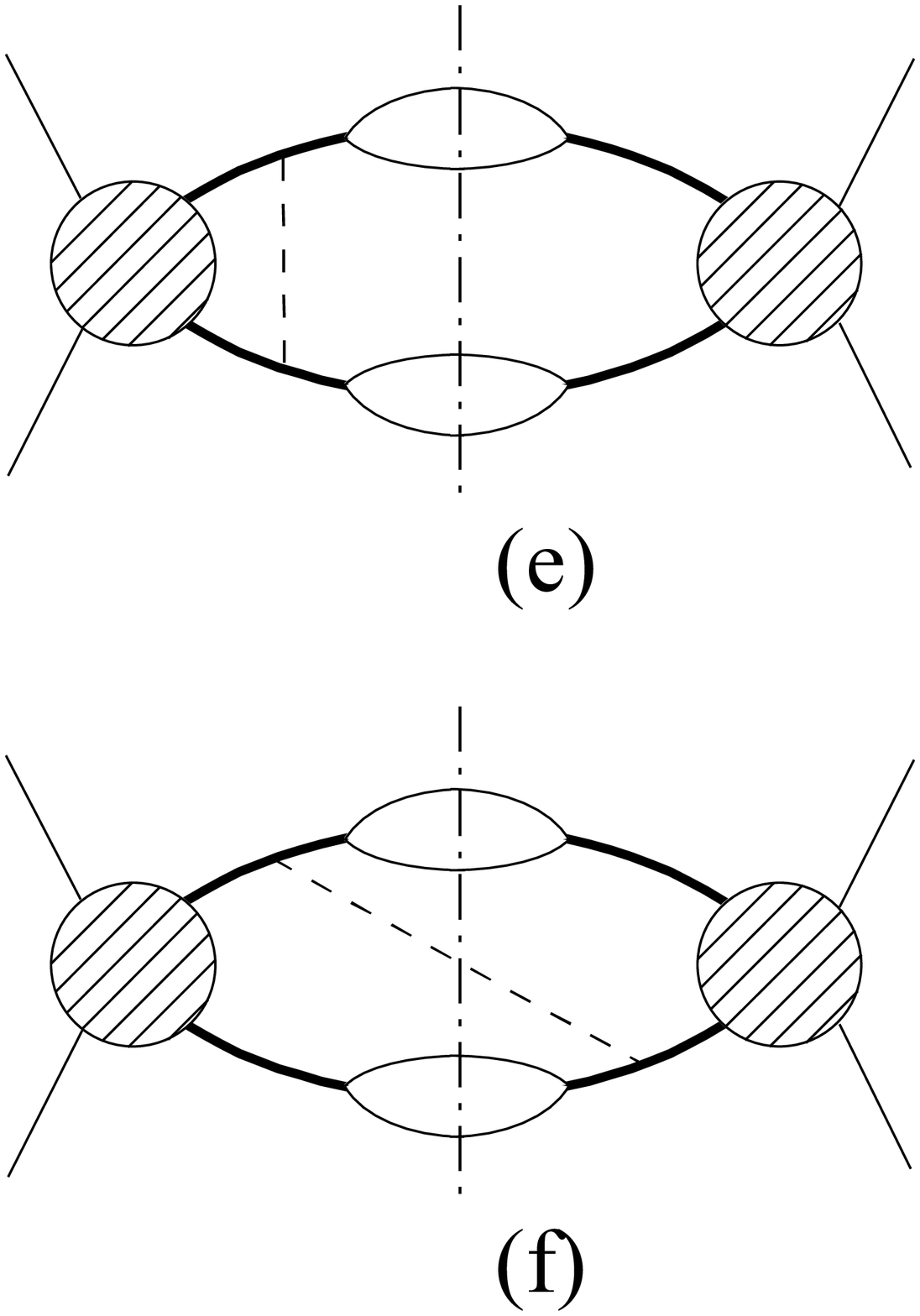} \hspace*{10pt}
       \epsfxsize=77pt \epsfbox{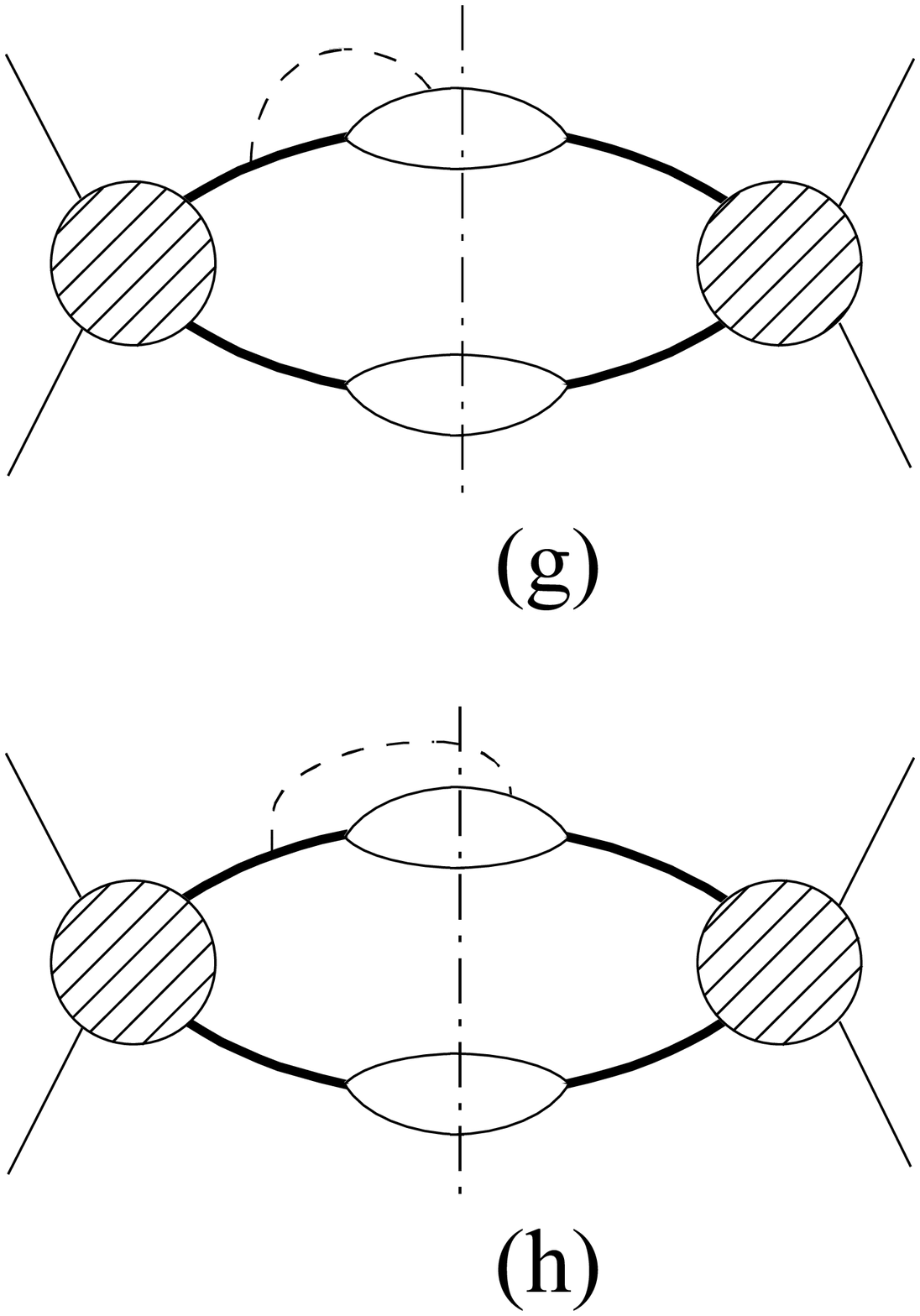} \hspace*{10pt}
       \epsfxsize=77pt \epsfbox{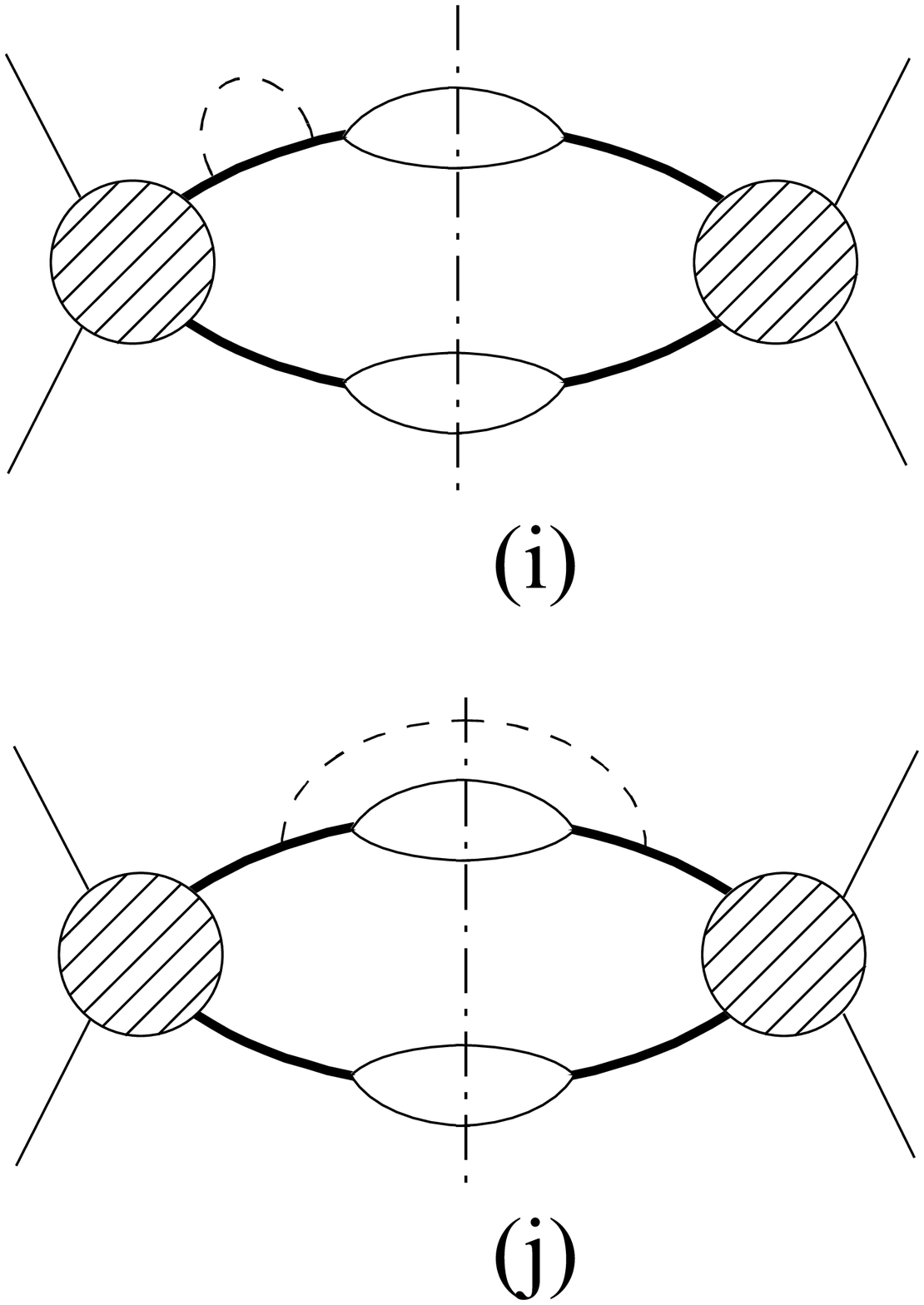} }
\caption{\small Examples for factorizable and (quasi)
non-factorizable corrections to the W-pair production which have
to be taken into consideration in the NLO approximation. (The
dashed lines denote the massive bosons or the photons. The thin
and thick lines represent the initial/final fermions and the
W-bosons. The vertical dot-dashed lines mean the cut of unitarity.
The hatched areas denote the corresponding lowest-order Green
functions.)} \label{fig:1}
\end{figure*}
\begin{figure*}
\hbox{ \hspace*{25pt}
       \epsfxsize=77pt \epsfbox{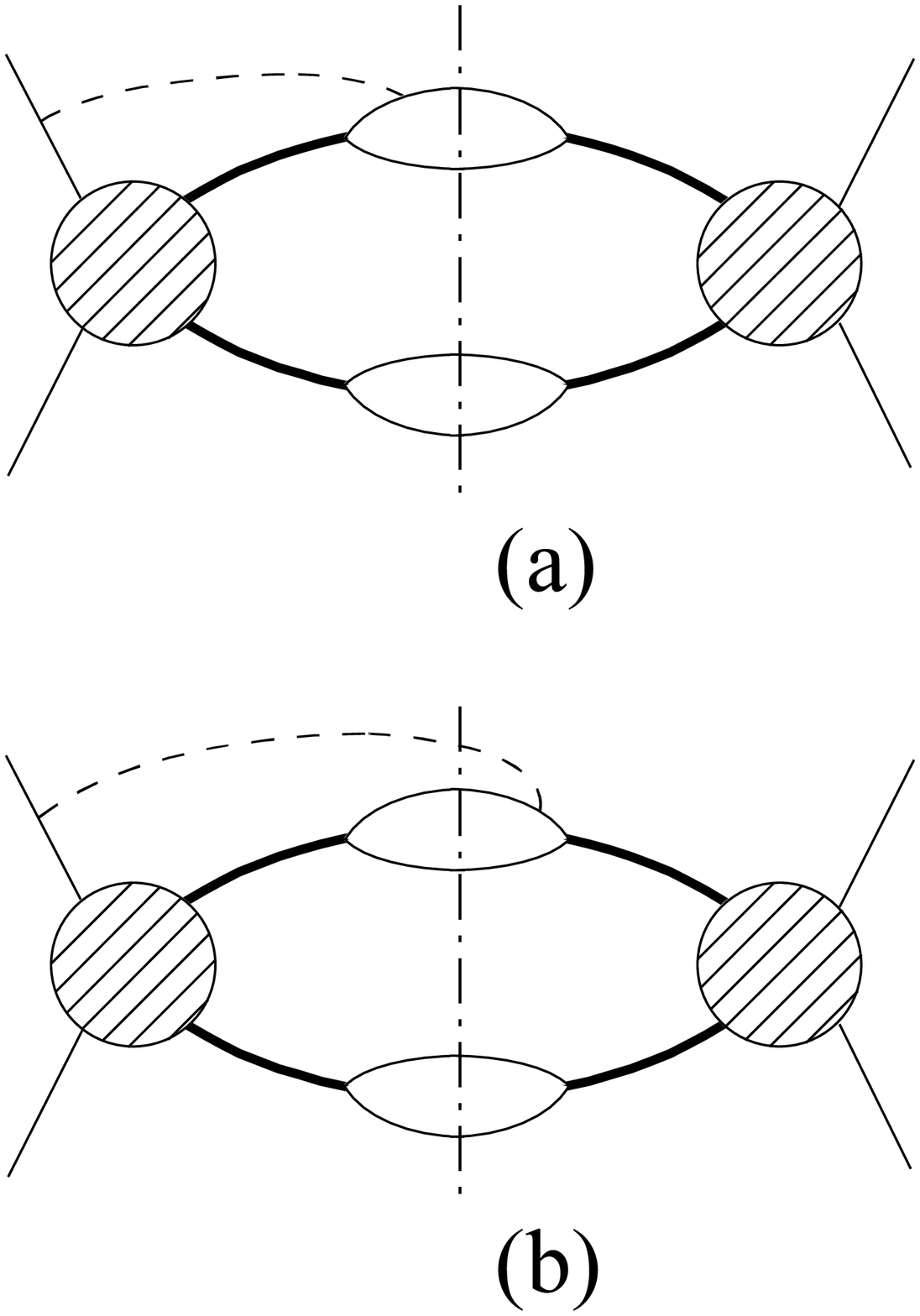} \hspace*{20pt}
       \epsfxsize=77pt \epsfbox{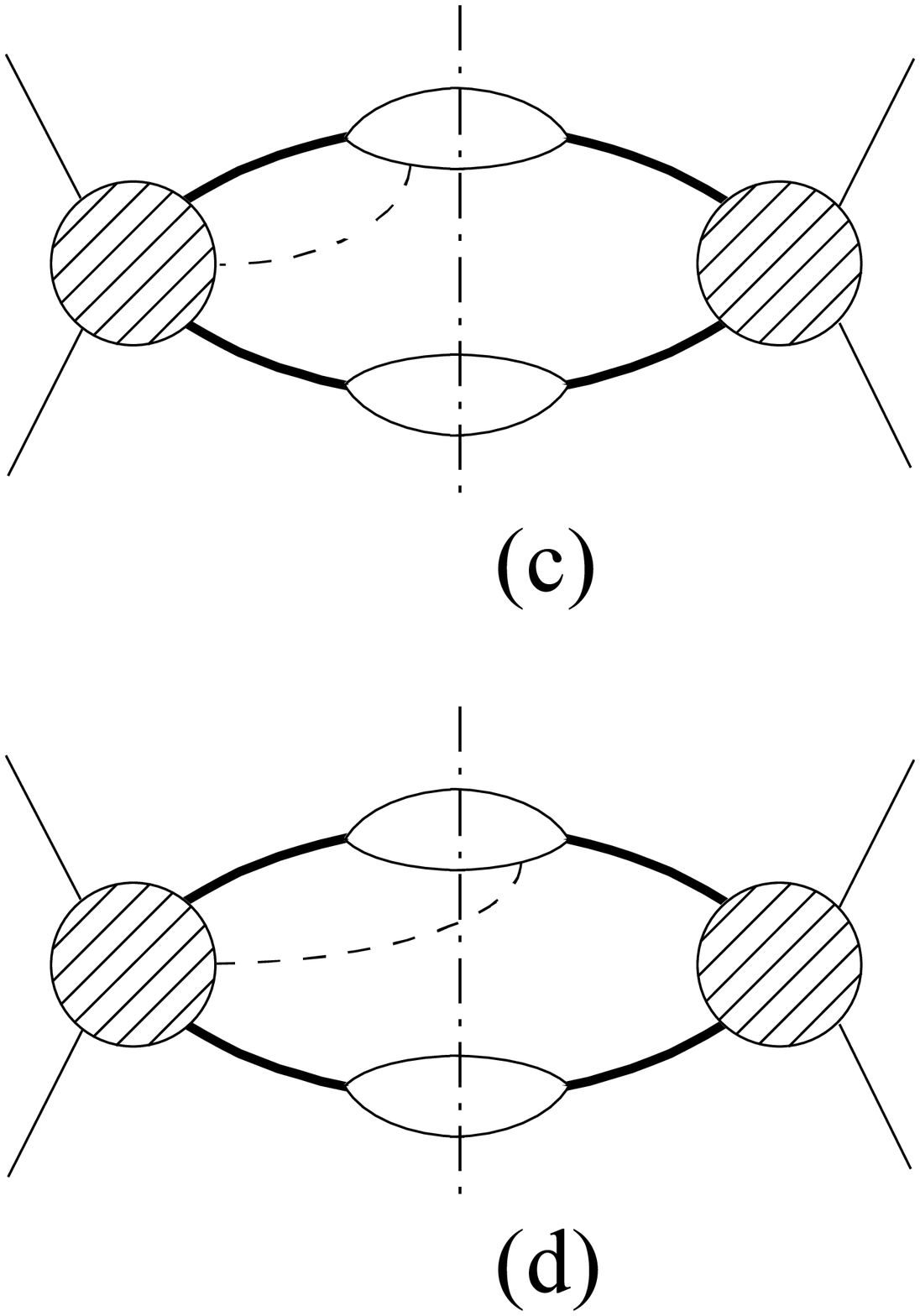} \hspace*{20pt}
       \epsfxsize=77pt \epsfbox{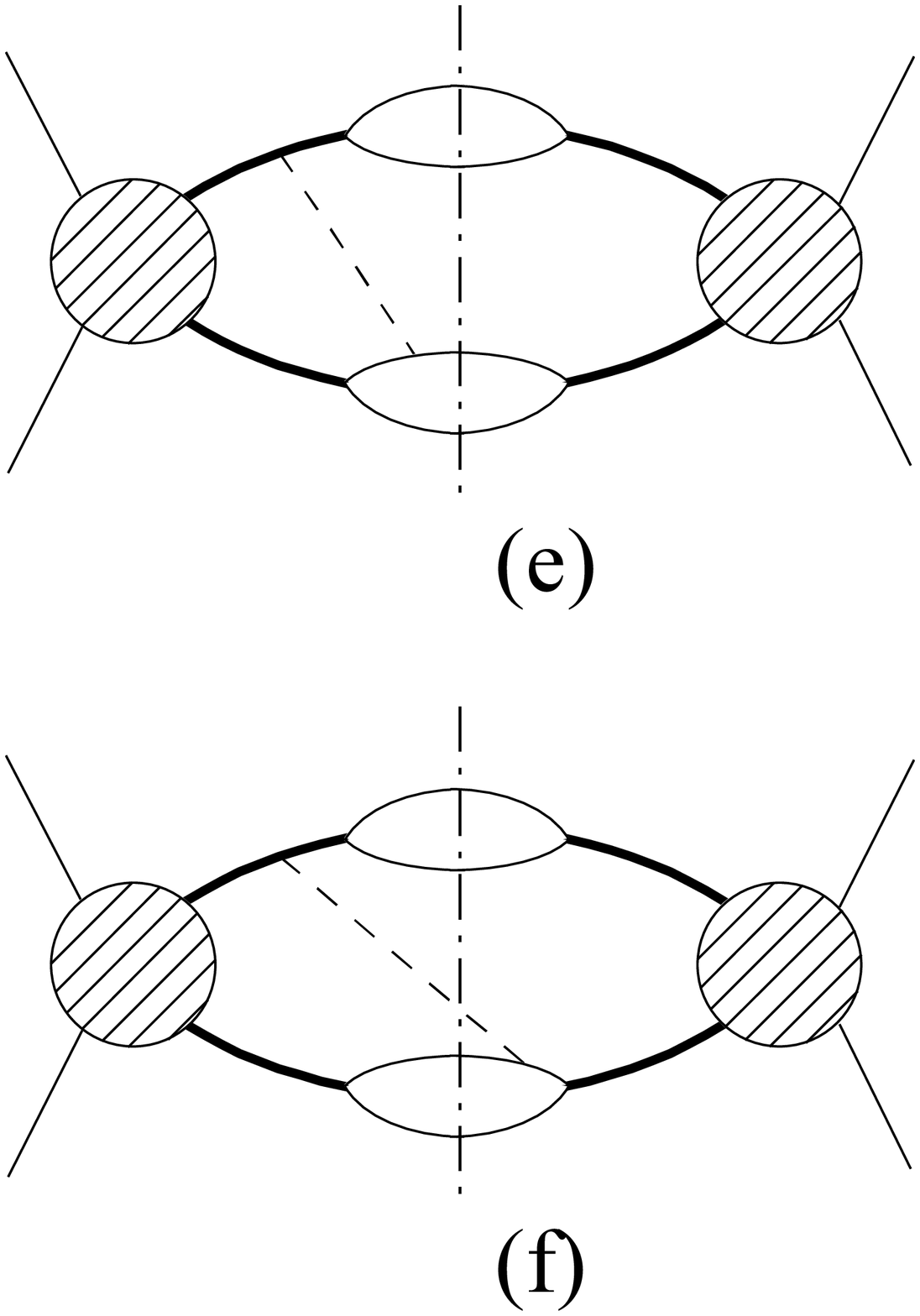} \hspace*{20pt}
       \epsfxsize=77pt \epsfbox{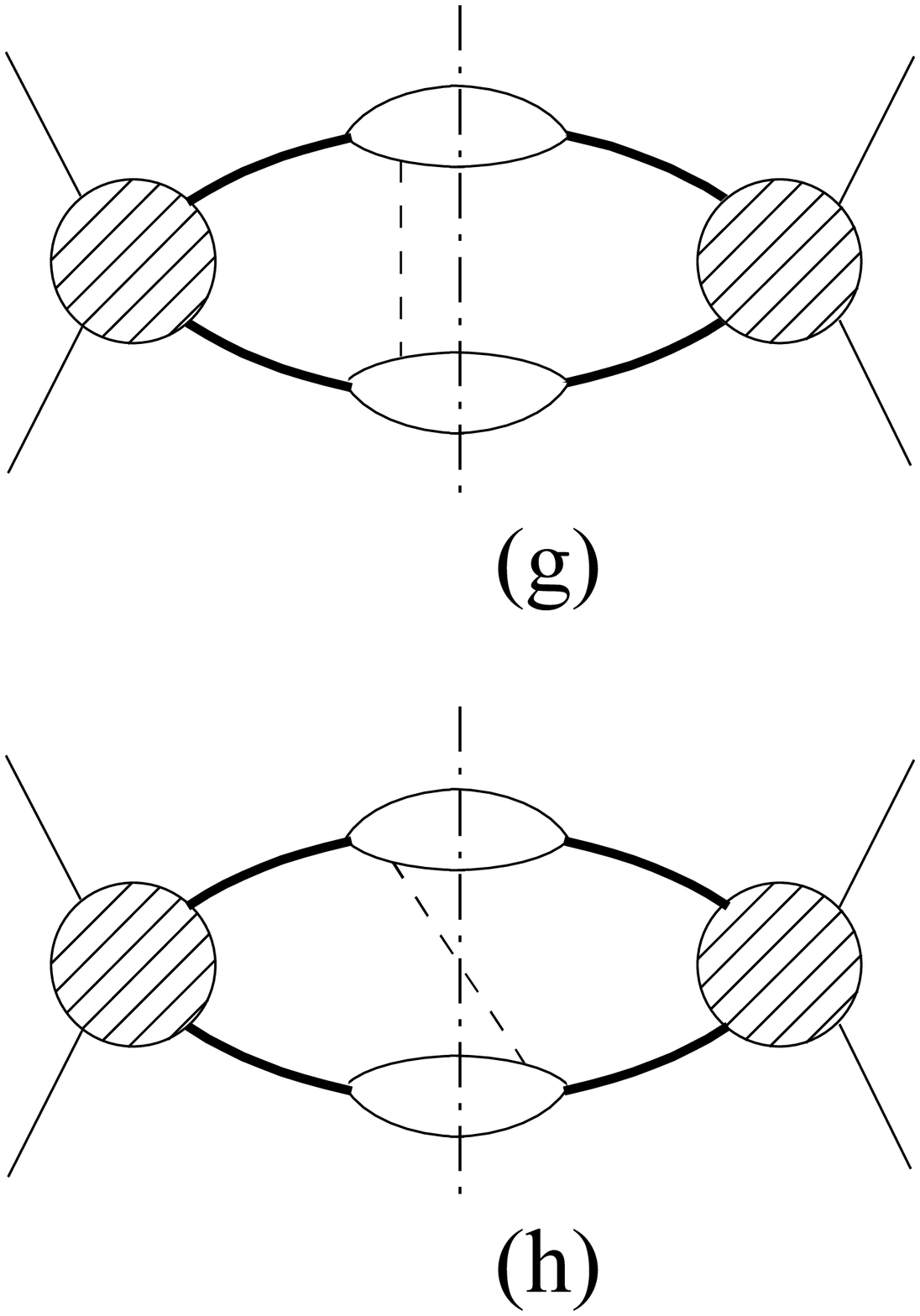} }
\caption{\small Examples for manifestly non-factorizable
corrections to the W-pair production. } \label{fig:2}
\end{figure*}

The above discussion leads us to the following resultant formulas
for the cross-section:
\begin{equation}\label{6}
\sigma (s) = \int\limits_{0}^1 \d z \> \phi (z;s) \> \hat\sigma
(zs)\,, \qquad \hat\sigma (s) = \int\limits_0^s{\d s_+
\!\!\int\limits_0^{(\sqrt s\;- \sqrt {s_+})^2} {\!\!\d s_-
\>\hat\sigma_0(s\,;s_+,s_-)}} .
\end{equation}
Here $\sigma(s)$ is the experimentally measured total
cross-section at the center-of-mass energy squared $s$. The $\phi
(z;s)$ is the ``flux'' function describing the contributions of
the initial-state and final-state photon radiations with large IR
and collinear logarithms. Quantity $\hat\sigma(s)$ is the hard
scattering cross-section at the reduced center-of-mass energy
squared (see Refs.~\cite{LEP2,NuovoCim,B-P} and references
therein). It should be noted that $\hat\sigma(s)$ contributes to
$\sigma(s)$ as distribution, because the $\hat\sigma(s)$ is
smeared by the flux function over the allowed range of kinematic
variable ($\phi (z;s)$ is peaked at $z=1$ and has a tail until a
cut at lower values of $z$). The second formula in (6) represents
$\hat\sigma(s)$ in form with the explicit phase-volume integration
(over the virtualities of the unstable particles, which are the
invariant-masses of the corresponding final states). The
$\hat\sigma_0(s)$ is the quantity that we have discussed above. It
reads as follows:
\begin{eqnarray}\label{7}
&\hat\sigma_0(s\,;s_+,s_-) = \hat\sigma_0^{\mbox{\scriptsize
fermion-loop-scheme}}(s\,;s_+,s_-)
\times(1+\delta^{\,\mbox{\scriptsize anom}})&
\nonumber\\[0.7\baselineskip]
& + \hat\sigma _0^{\mbox{\scriptsize {CC03-on-shell,
boson-one-loop+real-photon}}}(s\,;M_+,M_-) \times
\prod\limits_{\kappa=\pm}\delta(s_{\kappa} - M_{\kappa}^2)\times
B\!R_{\kappa}^{\;\mbox{\scriptsize tree}}&
\nonumber\\[0.4\baselineskip]
& + \hat\sigma_0^{\mbox{\scriptsize
CC03-on-shell,tree}}(s\,;M_+,M_-) \times \prod\limits_{\kappa=\pm}
\delta(s_{\kappa} - M_{\kappa}^2)\times
B\!R_{\kappa}^{\;\mbox{\scriptsize
{tree/boson-one-loop+real-photon}}} \;.&
\end{eqnarray}%
Here the first term (without the $\delta^{\,\mbox{\scriptsize
anom}}$) represents the result of the conventional fermion-loop
scheme. All other terms describe the corrections. Quantities
$B\!R_{\pm}^{\;\mbox{\scriptsize tree}}$ mean the lowest-order
on-shell branchings of the $\mbox{W}^{\pm}$-bosons. In the third
term one of $B\!R$'s is determined as the sum of the bosonic
one-loop correction to the partial width of W-boson and the width
with one real photon, divided by the total Born width. (In fact,
the third term includes the sum of four subterms, with the
modified $B\!R$ for one of two unstable particle.) All the above
ingredients are IR-finite and gauge-invariant. Moreover, except
for $\Im\Sigma_2(M^2)$ in $\delta^{\,\mbox{\scriptsize anom}}$,
all of them have already been calculated
\cite{W-pair-on-shell,W-decay}.

\medskip

\noindent $\mbox{\bf 4}$. In the above discussion we have omitted
the corrections caused by the gluon exchanges. However, they
contribute to the real processes and, so, they have to be taken
into consideration. The gluon exchanges occur between the
final-state quarks, between the final-state quarks and W-bosons at
mediating the quark-loops, and between the W-bosons at mediating
the quark-loops. Examples of the corresponding configurations are
shown in Fig.1c-j and Fig.2c-h with a modified specification.
Namely, under the dashed lines we imply now a group of  gluons
(two or more), and in the points of intersection between the
dashed lines and the W-boson lines, or the vertices, we imply the
presence of the fermionic loop. Correspondingly, these points of
intersection are supplied with the additional factor $\alpha$. (We
do not take into account the strong coupling constant $\alpha_s$,
assuming that it is of order $O(1)$. The latter assumption means
that we imply the soft-gluon exchanges. The hard-gluon exchanges
are additionally suppressed by the smallness of $\alpha_s$.)

In order to avoid the collinear and IR divergences we consider the
gluons at the intermediate stage to be massive (it is enough to
make the groups of the gluons to be effectively massive).
Simultaneously, this permits to delete from the list of diagrams
that contribute into the NLO the following configurations:
Fig.1c-f and Fig.1i,j as including the additional $\alpha^2$,
Fig.2c-f as suppressed by one additional $\alpha$ and switching
off the propagator squared, and Fig.2g,h as suppressed by
switching off the two unstable propagator squared. All the
above-mentioned configurations considered with the new
specification make contributions to the NNLO.

So, the configurations which are responsible for the phenomenon of
color reconnection make contributions beyond the NLO
approximation. The only configurations surviving in the NLO are
those which contribute to the decay factors of W-bosons. Among the
listed above configurations they are of Fig.1g,h. They have to be
taken into consideration while calculating the corresponding
$B\!R$'s in formula (7).

\medskip

\noindent $\mbox{\bf 5}$. The above results may be easily
generalized to the cases of the differential cross-sections. What
one must do in these cases is to replace the expressions that
stand behind the above-discussed diagrams of unitarity by the
corresponding amplitudes squared with saving the necessary
integrations (the configurations remain the same). Then, in the
case of the pure angular distributions, the only modification in
formulas (6) and (7) consists in replacing the cross-sections (and
branchings, if necessary) by the corresponding angular
distributions. The generalization to the cases of the
invariant-mass distributions is possible, as well, although it is
not so straightforward. Really, it should necessarily involve a
serious modification in the structure of the integrals in (6).
Namely, at least one of the two phase-volume integrals must
disappear in the second formula in (6). As a result, the
convolution becomes directly acting on the corresponding
hard-scattering distribution. This rather sharp modification may
have nontrivial consequences. In particular, it may result in the
change for the worse of the convergence properties of the series
in powers of $\alpha$ in the modified PT. In practice exactly this
change seems to have place, because the manifestly
non-factorizable corrections become non-negligible in the case of
the invariant-mass distributions \cite{Non}. Nevertheless, in view
of the sharp decreasing of the effect with the increasing $s$
\cite{Dittmaier,Non}, the property of convergence seems to rapidly
be improving with the energy increasing. Since the invariant-mass
distributions are not important for the practical usage, at least
near the threshold, the above-discussed effect should not have
important consequences.

In the cases of the total cross-section and the pure angular
distributions the property of convergence of the modified PT
remains quite satisfactory almost everywhere, except for the range
of the direct vicinity to the threshold which is numerically less
than the width of W-boson. This observation follows from the
well-known property of closeness in this range of the on-shell
CC03 total cross-section to the off-shell CC03 total cross-section
\cite{LEP2}. Actually, the difference between these two
cross-sections is a characteristic of the accuracy of the
description of the W-pair production in the leading-order
approximation. The numerical accuracy of the description within
the NLO approximation is controlled by this difference, as well.
The more detailed discussion of this problem is the subject of the
forthcoming paper.

\medskip

\noindent {\it Acknowledgements.} The author is grateful to
D.Yu.Bardin for support and helpful discussions.


\end{document}